\documentclass[tightenlines,aps,
twocolumn,
showpacs,nofootinbib]{revtex4}

\usepackage{psfig,float}
\usepackage{amsmath,amsfonts,graphicx,bm}

\newcommand{\ep}{\epsilon}
\newcommand{\be}{\begin{equation}}
\newcommand{\ee}{\end{equation}}
\newcommand{\ba}{\begin{eqnarray}}
\newcommand{\ea}{\end{eqnarray}}

\begin{document}

\begin{titlepage}

\begin{flushright}
\vbox{
\begin{tabular}{l}
FERMILAB-PUB-06-059-T\\
UH-511-1083-06\\
MADPH-06-1256
\end{tabular}
}
\end{flushright}

\title{
The $W$ boson production cross section at the LHC through ${\cal O}(\alpha_s^2)$
}

\author{Kirill Melnikov
        \thanks{e-mail: kirill@phys.hawaii.edu}}
\affiliation{Department of Physics and Astronomy,
          University of Hawaii,\\ 2505 Correa Rd. Honolulu, HI 96822}  
\author{Frank Petriello\thanks{frankjp@phys.wisc.edu}}
\affiliation{
Department of Physics, University of Wisconsin, Madison, WI  53706\\
and\\
Fermi National Laboratory, P.O. Box 500, MS 106, Batavia, IL 60510
} 

\begin{abstract}
We compute the ${\cal O}(\alpha_s^2)$ QCD corrections to  the 
fully differential cross-section $pp \to W X \to l  \nu  X$, 
retaining all effects from spin correlations.
The knowledge of these corrections makes it possible to calculate with high precision the 
$W$ boson production rate and acceptance at the LHC, subject to realistic cuts 
on the lepton and missing energy distributions.  For certain choices of cuts 
we find large corrections when going from next-to-leading order (NLO) 
to next-to-next-to-leading order (NNLO) in perturbation theory.  These corrections are significantly larger than those obtained by 
parton-shower event generators merged with NLO calculations.  
Our calculation may be used to assess and significantly reduce the QCD uncertainties in the  many studies of $W$ boson 
production planned at the LHC.
\end{abstract}

\maketitle

\thispagestyle{empty}
\end{titlepage}

Production of electroweak gauge bosons is a vital component 
of the hadron collider physics program.  The large production rates for this 
channel at the LHC will facilitate several important precision measurements.
LHC experiments plan to determine the $W$ boson mass and width with 
errors of $\delta M_W \sim 15~{\rm MeV}$ and
$\delta \Gamma_W \sim 50~{\rm MeV}$, respectively~\cite{wmeas}.  The Weinberg angle
$\sin \theta_W$ can be extracted from
the forward-backward asymmetry of the lepton pair 
in  $pp \to Z \to e^+e^-$ with a precision of $1 \times 10^{-4}$.
The precision possible in these channels at high luminosities 
makes these measurements competitive with LEP results.
Searching for deviations from predictions in di-lepton events with 
large invariant mass, missing energy, or transverse momentum probes extensions 
of the Standard Model which contain new gauge bosons.

In addition to its high rate, electroweak gauge boson production has a simple, distinct experimental 
signature.  This also makes it a useful process for calibrating and monitoring machine and detector performance.
$Z$ and $W$ production can be used to determine and
monitor the hadronic and partonic luminosities at 
the LHC~\cite{lum}.  This requires a theoretical prediction for the 
cross section to the highest possible precision, since this 
error propagates into all other measurements through the luminosity uncertainty.
Determination of the LHC luminosity to 1\% accuracy is the ultimate goal of this procedure~\cite{lum}.  
This sets the precision required for theoretical predictions.

When the desired precision on the production cross section is at 
the few percent level, many subtle effects must be included.  Both 
${\cal O}(\alpha)$ electroweak
effects and ${\cal O}(\alpha_s^2)$ QCD effects must be calculated.
The electroweak corrections to $pp \to W \to l \nu$
were computed in~\cite{ew1}, where the importance of final state 
photon radiation in the $W$ decay was observed.  NLO 
computations of the QCD corrections to electroweak gauge boson production were 
first obtained in the late seventies~\cite{nlo}.  The $W$ boson momentum 
distribution was investigated via resummation techniques in \cite{ptW}.
Currently, the NNLO QCD corrections 
are known for both the inclusive production cross section~\cite{DY} 
and for the gauge boson rapidity distributions~\cite{Anastasiou:2003ds}. 
The NNLO corrections are typically in the few 
percent range at the LHC, and must be included in both precision 
electroweak studies and the luminosity determination.

Existing calculations of the NNLO QCD corrections to this process do not include 
all effects needed for a percent level theoretical prediction.  Phenomenological 
applications of $Z$ and $W$ production require significant cuts on the 
phase-space of the final state leptons.  For example, all LHC experiments will impose constraints on the transverse 
momenta and rapidities of the final state charged leptons.  Cuts on the missing energy will also 
be employed to identify the neutrino from $W$ decay.  Calculations that treat the leptons inclusively 
are therefore not fully realistic.  They can be used to make estimates, but they are 
not sufficient for precision measurements.

The calculation of the full NNLO QCD corrections is complicated by the 
spin-one nature of the gauge bosons.  If they were spin-zero bosons, 
fully differential results could be obtained from Ref.~\cite{Anastasiou:2003ds}, where the rapidity 
distributions for $Z$ and $W$ bosons were computed through NNLO.  That result could be 
combined with the known double differential distribution in transverse momentum
and rapidity~\cite{keith} to fully determine the gauge boson kinematics.  The decay of spin-zero 
bosons in their rest frame is isotropic, 
and the final state distribution of interest could be 
obtained by assuming a flat decay distribution in the gauge boson rest frame.  There would 
be no correlation between the production and decay of the boson.  
However, since the $Z$ and $W$ are spin-one bosons, 
the $Z \to l^+ l^-$ and $W \to l \nu$ 
decays are not isotropic, and 
there are "spin correlations" 
between the production and decay channels.  The importance of spin 
correlations for the production of gauge bosons at the LHC was
recently emphasized in~\cite{Frixione:2004us}.  

The computation of the differential cross section for $pp \to WX \to l  \nu X$ through NNLO in QCD 
is a difficult theoretical challenge.  While techniques for 
performing differential NLO calculations have been known for many years~\cite{nlogen}, 
the corresponding technology for obtaining NNLO results is still in its infancy.
In a recent series of papers~\cite{method}, we have developed a method for performing these 
calculations.  This technique features an automated extraction of infrared singularities 
from the real radiation matrix elements and a numerical cancellation of these divergences with 
the virtual corrections.  We describe below the application of this method to the 
computation of the $W$ production cross section with all spin correlations included.

We compute the partonic cross sections $i  j \to l \nu X$ as perturbative expansions 
in the strong coupling constant $\alpha_s$.  We specialize here to $W^-$ production.  
At LO, the $W^-$ boson is produced in the collision of an up-type antiquark and a down-type quark.
At NLO, gluon-quark and gluon-antiquark scatterings also contribute.  A variety 
of partonic processes contribute at NNLO.  These have been enumerated in great 
detail in~\cite{DY}.  In our discussion below we use $\bar u  d \to W^-X \to e  \bar \nu X$ 
as an example, since it contains all the complexities present in the full calculation.  
All partonic channels have been included in our result.

There are three distinct contributions contained in the $\bar u  d$ initiated process: the 
two-loop virtual corrections, the one-loop virtual corrections to single gluon emission, 
and tree-level double-real radiation processes with two additional partons in the final state.  These must be combined 
in the presence of an infrared-safe measurement function to produce a finite result.  We 
use dimensional regularization to regulate all ultraviolet, soft, and collinear divergences.

The two-loop virtual corrections to the $\bar u d$ process 
are straightforward to compute.  They are very similar
to the ${\cal O}(\alpha_s^2)$ corrections to 
the quark form factor studied in \cite{gonzalves};
however, 
the $W$ production calculation must include 
the two-loop corrections to non-singlet axial current.  Care 
must be taken to define this correctly in $d = 4-2\ep$ dimensions; we discuss this 
issue further below.  We use the implementation of the Laporta algorithm~\cite{Laporta} 
described in Ref.~\cite{AIR} to reduce all required two-loop integrals to a minimal 
set of master integrals.  The master integrals needed for this 
computation are well known~\cite{gonzalves}.

We obtain the one-loop correction to the single gluon emission process 
$\bar u  d \to e   \bar \nu_e  g$ using a combination of two methods. 
We first use the Laporta algorithm to express all one-loop 
Feynman integrals relevant for this process through master integrals. 
The master integrals must be integrated over the final state phase-space 
subject to the kinematic constraints under consideration.  It is not possible to perform 
this integration analytically, since we want an expression valid for arbitrary cuts.  
Numerical integration is also not straightforward because of soft and 
collinear singularities.  We employ the method developed in~\cite{method} to 
extract the singularities in a constraint-independent way as poles in $\ep$ before 
integrating over 
the phase-space numerically. This technique maps the final state phase-space 
onto the unit hypercube and uses iterated sector decomposition~\cite{sector} to extract all 
soft and collinear singularities.

We use essentially the same algorithm to compute the double-real radiation corrections.  A 
detailed description of this method, which studies in detail both the one and two parton 
emission corrections, can be found in~\cite{method}.

We now discuss a few new features of this calculation, first explaining how we treat 
the 
axial current in $d$-dimensions.  
This issue arises from Dirac structures 
of the form ${\rm Tr}_{\rm H}[\Gamma^{(1)} \gamma_5]
{\rm Tr}_{\rm L}[ \Gamma^{(2)} \gamma_5]$, where $\Gamma^{(1,2)}$ denote 
generic products of Dirac matrices and  ${\rm Tr}_{\rm H,L}$
refer to traces over hadronic and leptonic degrees of freedom, respectively.  These 
traces do not vanish when the final state phase-space is sufficiently constrained.  
A consistent extension of the axial current to $d$-dimensions is 
given in~\cite{larin}.  It utilizes an anti-commuting $\gamma_5$ and 
contains additional renormalizations relative to 
the vector current in order to maintain the Ward identities.  We use 
this prescription in our calculation.


Even after all three components of the hard-scattering cross section are combined, 
collinear counterterms are needed to remove initial state collinear singularities.
In~\cite{method} these collinear counterterms were treated analytically.  Such 
an approach is not sufficiently flexible to handle cuts on the $W$ decay products.  
However, it is straightforward to extend the numerical approach used for the other 
NNLO components in~\cite{method} to obtain the desired results.  

We have essentially two checks on our calculation.  
First, considering  different  cuts on the
electron transverse momentum and rapidity
as well as on the missing energy, we verify
cancellation of the divergences in the $W^-$ production cross section.
Because the divergences start at $1/\ep^{4}$ at NNLO, the
cancellation of all divergences through $1/\ep$ provides a stringent 
check on the calculation.  We also check that the vector and axial contributions 
are separately finite, as required.  A second check is obtained by 
integrating fully over the final state phase-space and comparing against 
known results for the inclusive cross section.  We find excellent agreement 
with the results of~\cite{DY} for all partonic channels.

We now discuss the results of our calculation.  
We first present the input parameters.
We use the MRST parton distribution functions~\cite{mrst} at the appropriate order 
in $\alpha_s$.  We use $m_W = 80.451~{\rm GeV}$ and work in the narrow width approximation, 
although this restriction can be easily removed.  We set $|V_{\rm ud}| = 0.974$, $|V_{\rm us}| = |V_{\rm cd}| = 0.219$, 
and $|V_{\rm cs}| = 
0.996$, and obtain $|V_{\rm ub}|$ and $|V_{\rm cb}|$ from unitarity of the CKM matrix.  We neglect contributions 
from the top quark; these have been shown to be small in the inclusive cross section~\cite{DY}.  
For 
electroweak input parameters, we use $\sin^2 \theta_W = 0.2216$, $\alpha_{\rm QED}(m_Z) = 1/128$, 
and ${\rm Br}(W \to e \nu) = 0.1068$.  
We set the factorization and renormalization scales to a common value, $\mu_r=\mu_f=\mu$, and employ 
various choices of $\mu$ in our numerical study.


We find that NNLO corrections depend on the cuts and can change 
rapidly from very small to fairly substantial.
We consider cuts of the form 
\begin{equation}
p_\perp^e > p_\perp^{e, \rm min},\;\;\; |\eta^e| < 2.5,\;\;
E_\perp^{\rm miss} > 20~{\rm GeV},
\label{eq3}
\end{equation}
and use the values $p_\perp^{e, \rm min}=20,30,40,50$ GeV.  The choices 
$p_\perp^{e, \rm min}=20$ and 40 GeV were considered in the study of~\cite{Frixione:2004us}.
$p_\perp^{e, \rm min}=20$ GeV is similar to cuts that will be employed by the ATLAS and CMS 
collaborations, while $p_\perp^{e, \rm min}=40$ GeV was chosen in 
\cite{Frixione:2004us}
to illustrate the potential 
sensitivity of the QCD radiative corrections to experimental cuts.  We first present results 
for the lepton invariant mass distribution for on-shell $W^-$ production in Table~\ref{table1}, to give a 
feeling for the magnitude of the cross section for each set of cuts.  The numerical precision for all NNLO numbers 
is 1\% or better.  We note that the row labeled "Inc" denotes the fully inclusive cross section.

\begin{tiny}
\begin{table}[htbp]
\begin{center}
\begin{tabular}{|c|c|c|c|}
\hline\hline
$p_\perp^{e, \rm min}$ & LO & NLO & NNLO \\ \hline\hline
Inc & 11.70,13.74,15.65 & 16.31,16.82,17.30 &16.31, 16.40, 16.50 \\ \hline
20 & 5.85,6.96,8.01 & 7.94,8.21,8.46 & 8.10,8.07,8.10 \\ \hline 
30 & 4.305, 5.12,5.89 & 6.18,6.36,6.54 & 6.18,6.17,6.22 \\ \hline 
40 & 0.628,0.746,0.859 & 2.07,2.10,2.11 & 2.62,2.54,2.50 \\ \hline
50 & 0,0,0 & 0.509,0.497,0.480 & 0.697,0.651,0.639 \\ \hline\hline
\end{tabular}
\caption{\label{table1} The lepton invariant mass distribution $d\sigma/dM^2$,
$M=m_W$, for on-shell
$W$ production in the reaction $pp \rightarrow W^-X \rightarrow e^- \bar \nu W$, in ${\rm pb/GeV}^2$, for
various choices of $p_\perp^{e, \rm min}$, GeV
and $\mu = m_W/2,m_W,2m_W$.}
\vspace{-0.1cm}
\end{center}
\end{table}
\end{tiny}

\vspace*{-0.3cm}
There are a few things to notice about these numbers.  First, at LO, there is no additional hadronic 
radiation in the final state for the lepton and neutrino to recoil against, so the transverse 
momentum is restricted to $p_\perp^{e}<m_W/2$.  The cross section is therefore very small for 
$p_\perp^{e, \rm min} = 40$ GeV, and vanishes for $p_\perp^{e, \rm min} = 50$ GeV.  This restriction is lifted at NLO 
when there is an additional parton for the $W$ to recoil against.  Very near this boundary, the width of the 
$W$ can be an important effect.  It will induce a (tiny) cross section for $p_\perp^{e, \rm min} = 50$ GeV at LO, 
and it will shift the result for $p_\perp^{e, \rm min} = 40$ GeV since this value is close to $m_W/2$.  
We have not included the $W$-width in our 
results.  However, we have checked using the results in~\cite{Frixione:2004us} that finite width effects change the acceptance by only 7\% 
at LO, and by less at NLO.  We are therefore confident that our discussion and conclusions are not affected by this omission.

Another feature to notice is that the corrections are large for higher choices of $p_\perp^{e, \rm min}$ and that the dependence on 
$p_\perp^{e, \rm min}$ is strong.
For example, for the scale choice $\mu=m_W$,
we observe a 22\% increase when going from 
LO to NLO inclusive cross section, 
followed by a decrease of 2.5\% when  NNLO corrections 
are included.  We obtain similar results 
for $p_\perp^{e, \rm min}=20,30$ GeV. 
However, the pattern of corrections is much different 
for $p_\perp^{e, \rm min}=40,50$ GeV; we find corrections of 18-27\% for $p_\perp^{e, \rm min}=40$ GeV and 
33-37\% for 50 GeV when going from NLO to NNLO, depending on the choice of scale.  

The remaining scale dependences can be seen from Table~\ref{table1}. 
We define the scale dependences of the cross section $\sigma_X$ 
as $\Delta \sigma_X = 2
({\rm Max}[\sigma_X(\mu)]-{\rm Min}[\sigma_X(\mu)])/({\rm Max}[\sigma_X(\mu)]+{\rm Min}[\sigma_X(\mu)])$, where we take the maximum and 
minimum values from among the three studied scale choices.
$\Delta \sigma_X$ therefore gives the scale variation 
uncertainty band for the observable $\sigma_X$. 
The scale dependence is reduced to the 
percent level or less for the inclusive case, and for the cuts $p_\perp^{e, \rm min}=20,30$ GeV.  Moreover, the NNLO results 
lie within the NLO uncertainty bands.  This is not the case for the other two choices of $p_\perp^{e, \rm min}$; here, the scale 
dependence actually increases to the 5-8\% level at NNLO, 
and the NLO scale dependence completely underestimates 
the higher-order radiative corrections.  This is not completely unexpected, since for these values 
additional partons to recoil against only appear at NLO. 
The NNLO results therefore serve as the first radiative corrections 
for these $p_\perp^{e, \rm min}$ choices.  However, it indicates the care that must be taken when using the scale variation as a 
measure of the theoretical error.



\begin{tiny}
\begin{table}[htbp]
\begin{center}
\begin{tabular}{|c|c|c|}
\hline\hline
$p_\perp^{e, \rm min}$ (GeV) & $A({\rm NLO})$ & $A({\rm NNLO})$ \\ \hline\hline
20 & 0.487,0.488,0.489 & 0.497,0.492,0.491 \\ \hline
30 & 0.379,0.378,0.378 & 0.379,0.376,0.377 \\ \hline
40 & 0.127,0.125,0.122 & 0.161,0.155,0.152 \\ \hline
50 & 0.0312,0.0295,0.0277 & 0.0427,0.0397,0.0387 \\ \hline\hline
\end{tabular}
\caption{\label{table4}
Acceptances at NLO and NNLO
for various choices of $p_\perp^{e, \rm min}$ and $\mu=m_W/2,m_W,2m_W$.
}
\vspace{-0.1cm}
\end{center}
\end{table}
\end{tiny}

\vspace*{-0.3cm}
Another important quantity to study is the experimental acceptance, defined as the ratio of 
the cross section after cuts over the inclusive cross section.
We present the acceptances at NLO and NNLO in Table~\ref{table4}.
We again note that for the choices $p_\perp^{e, \rm min}=40,50$ GeV, the NLO scale dependences completely
underestimate the NNLO corrections.
The NNLO shifts in acceptances are very large for $p_\perp^{e, \rm min} = 40,50$ GeV, reaching 25\% for
40 GeV and 40\% for 50 GeV.  For the other choices of
cuts, the NNLO acceptances are identical to the NLO ones within numerical errors, indicating stabilization of the perturbative expansion.

The cross sections and acceptances for the transverse momentum cuts $p_\perp^{e, \rm min}=20,40$ GeV 
were recently studied in~\cite{Frixione:2004us}.  The primary tool used in that analysis was the 
Monte Carlo event generator MC@NLO, which consistently combines NLO corrections with the 
HERWIG parton shower~\cite{webber}.  In~\cite{Frixione:2004us}, MC@NLO is used to estimate 
the importance of QCD effects beyond those included in NLO calculations.  Differences between 
the cross sections and acceptances of a few percent are found when comparing NLO and MC@NLO 
for both $p_\perp^{e, \rm min}$ choices.  The authors of~\cite{Frixione:2004us} then conclude 
that higher order corrections beyond those in MC@NLO are unlikely to change significantly 
the results they find.

We believe that the few percent differences between MC@NLO and NLO cannot be used 
as an estimate of higher order corrections. 
This is because few percent 
shifts coming from hard emissions generically occur in process-dependent 
radiative corrections that cannot be described by parton showers.
We can support this assertion with the following observations.  It 
follows from Table 2 in~\cite{Frixione:2004us} that adding the parton shower to the LO 
cross section fails to properly predict the NLO cross section.  For $p_\perp^{e, \rm min}=20$ GeV, adding 
the parton shower to the LO result {\it decreases} the cross section by 8\%, while the 
NLO correction increases it by 4\%.  While both HERWIG and NLO corrections 
increase the LO result for $p_\perp^{e, \rm min} = 40$ GeV, the magnitude of the shifts differ by 45\% relative to the LO cross section.  In addition, the 
NNLO corrections to the cross-sections presented in Table~\ref{table1} differ 
significantly from the estimate of these corrections in \cite{Frixione:2004us}.
We conclude that the results for 
the cross sections and 
acceptances obtained in~\cite{Frixione:2004us} cannot be 
used if few percent precision is required.  This is particularly true for 
cuts where hard gluon emissions are expected to be large, such as 
for $p_\perp^{e, \rm min}=40,50$ GeV. We note
that since the corrections are large for $p_{\perp}^{e, \rm min}=50$ GeV, which
is well above the LO kinematic boundary for the electron transverse
momentum  at $m_W/2$, our results are not caused by
large logarithms which spoil the perturbative expansion.

In this Letter we report on the computation of the NNLO QCD corrections to the 
fully differential cross section $pp \to WX \to l \nu X$ at the LHC, with all
spin correlations included.  For inclusive enough cuts,
this calculation provides the percent-level theoretical accuracy 
needed for the $W$ cross section when realistic experimental cuts are imposed 
on the final state leptons.  We find that the QCD corrections exhibit significant 
dependence on the lepton minimum transverse momentum.  For high values of this cut, the corrections may be 
very different than the inclusive NNLO results.  
Our calculation can be easily extended to include $Z$ production, finite 
width effects, and $p \bar p$ collisions.  The last case is particularly interesting 
because of its importance for the Tevaton Run II physics program.  These extensions 
will be discussed in detail in a forthcoming publication.

{\bf Acknowledgments} 
K.M. is supported in part by the DOE grant DE-FG03-94ER-40833, Outstanding 
Junior Investigator Award and by the Alfred P.~Sloan Foundation. 
F.P. is supported in part 
by the University of Wisconsin Research Committee 
with funds provided by the Wisconsin Alumni Foundation. 



\begin{thebibliography}{29}

\bibitem{wmeas}
For a review, see P.~M.~Nadolsky,
  AIP Conf.\ Proc.\  {\bf 753}, 158 (2005);
U.~Baur,
  arXiv:hep-ph/0511064.

\bibitem{lum}
M. Dittmar, F. Pauss and D. Z\"urcher, 
Phys. Rev. {\bf D56}, 7284 (1997);
V. A. Khoze {\it et al.},
 Eur. Phys. J. {\bf C19}, 313 (2001);
W.T. Giele, {\it et al.} 
arXiv:hep-ph/0104053.

\bibitem{ew1} 
S.~Dittmaier and M.~Kramer, Phys. Rev. {\bf D65}, 073007 (2002);
U.~Baur {\it et al.}, Phys. Rev. {\bf D59}, 013002 (1999).



\bibitem{nlo} 
G.~Altarelli {\it et al.}, Nucl. Phys. {\bf B157}, 461 (1979);
J.~Kubar-Andre {\it et al.}, Phys. Rev. {\bf D19}, 221 (1979);
K.~Harada {\it et al.}, Nucl.Phys. {\bf D155}, 169 (1979) 
[Err.-ibid: {\bf B165}, 545 (1980)]; P.~Aurenche and J.~Lindfors, 
Nucl. Phys. {\bf B185}, 274 (1981).

\bibitem{ptW} C.~Balazs, C.-P.~Yuan, Phys. Rev. {\bf D56}, 5558 (1997);
F.~Landry {\it et al.} Phys. Rev. {\bf D67}, 073016 (2003).

\bibitem{DY}
R. Hamberg, W.L. van Neerven and T. Matsuura, Nucl. Phys. {\bf B359}, 
343 (1991)
[Err.-ibid. {\bf B644}, 403 (2002)]; R.V.~Harlander and W.B.~Kilgore, 
Phys. Rev. Lett. {\bf 88}, 201801 (2002).


\bibitem{Anastasiou:2003ds}
C.~Anastasiou {\it et al.}, 
Phys. Rev. {\bf D69},
094008 (2004).

\bibitem{keith} 
R.~K.~Ellis {\it et al.}, Nucl. Phys. {\bf B211}, 106 (1983);P.~B.~Arnold and M.~H.~Reno, Nucl. Phys. {\bf B319}, 37 (1989); [Err.-ibid:
{\bf B330}, 284 (1990)].


\bibitem{Frixione:2004us}
  S.~Frixione and M.~L.~Mangano,
  JHEP {\bf 0405}, 056 (2004).


\bibitem{nlogen}
R.K.~Eliis {\it et al.}, Nucl. Phys. {\bf B178}, 421 (1981); 
W. T. Giele and E. W. N. Glover, Phys. Rev. {\bf D46}, 1980 (1992);
Z.~Kunszt and D.~E.~Soper, Phys. Rev. {\bf D46},  192 (1992);
S.~Frixione {\it et al.}, Nucl. Phys. {\bf B467}, 399 (1996);
S. Catani and M. H. Seymour, 
Nucl. Phys. {\bf B485} (1997), 291 [ Err.-ibid. {\bf B510}, 503 (1997)].


\bibitem{gonzalves}  R.~Gonzalves, Phys. Rev. {\bf D28}, 1542 (1983);
G.~Kramer {\it et al.}, Z. Phys.{\bf C34}, 497 (1987);
[Err.-ibid: {\bf C42}, 504 (1989)].



\bibitem{method} 
C.~Anastasiou, K.~Melnikov and F.~Petriello, Phys. Rev. {\bf D69}, 076010 
(2004); Phys. Rev. Lett. {\bf 93}, 032002 (2004); 
Phys. Rev. Lett. {\bf 93}, 262002 (2004);
Nucl. Phys. {\bf B724}, 197 (2005).

\bibitem{Laporta} S. Laporta, Int. J. Mod. Phys. {\bf A15}, 5087 (2000).

\bibitem{AIR} C.~Anastasiou {\it et al.}, JHEP {\bf 0407}, 046 (2004).

\bibitem{sector}
T. Binoth and G. Heinrich,  Nucl. Phys. {\bf B585}, 741 (2000); see also 
K.~Hepp, Comm. Math. Phys. {\bf 2}, 301 (1966);
A.~Denner {\it et al.}, Nucl. Phys. {\bf B479}, 495 (1996).

\bibitem{larin}
S.~Larin, Phys. Lett. {\bf B303}, 113 (1993).



\bibitem{mrst} A.D.~Martin {\it et al.}, Phys. Lett. {\bf B531}, 216 (2002).

\bibitem{webber} 
S.~Frixione and B.R.~Webber, JHEP {\bf 0206}, 029 (2002); 
S.~Frixione {\it et al.}, JHEP {\bf 0308}, 007 (2003).



\end{thebibliography}
\end{document}